\documentclass[]{spie}
\usepackage[dvips]{graphicx}

\title{Parrondo's games and the zipping algorithm}

\author{Pau Amengual and Ra\'{u}l Toral 
\skiplinehalf
Instituto Mediterr\'aneo de Estudios Avanzados, IMEDEA
(CSIC-UIB),\\ ed. Mateu Orfila, Campus UIB, E-07122 Palma de
Mallorca, Spain} 
\authorinfo{emails:pau@imedea.uib.es, raul@imedea.uib.es; http://www.imedea.uib.es/PhysDept}

\begin{document}
\maketitle
\begin{abstract}
We study the relation between the discrete--time version of the
flashing ratchet known as Parrondo's games and a compression
technique used very recently with thermal ratchets for evaluating
the transfer of information -- negentropy -- between the Brownian
particle and the source of fluctuations. We present some results
concerning different versions of Parrondo's games, showing all of
them a good qualitative agreement between the gain and the inverse  of the entropy.
\end{abstract}

\section{INTRODUCTION}
\label{sect:intro}  

In the last years, the microscopic field of Brownian particles has
received much attention due to new transport phenomena under
certain conditions. The main features involve breaking of spatial
inversion symmetry and a thermal bath that drives the system out
of equilibrium \cite{PR02} in order to obtain directed transport.
The former condition is accomplished through an asymmetric
potential, usually a ratchet--like potential. The latter can be achieved in different ways, either by a periodic or stochastic
forcing -- what it is known as \textit{pulsating ratchets}, or by
an additive driving force unbiased on average -- known as
\textit{tilting ratchets}. The \textit{flashing ratchet} model
corresponds to the class of pulsating ratchets and it consists on
switching on and off either periodically or stochastically a
ratchet potential. This model has been used recently for DNA
transport \cite{BH99} and separation of biological macromolecules
\cite{DA98}.

Recently, Arizmendi \textit{et. al} \cite{ASF03} have quantified
the transfer of information -- negentropy -- between the Brownian
particle and the nonequilibrium source of fluctuations acting on
it. They coded the particle motion into a binary string of $0$'s
and $1$'s, and then zipped  the resulting binary file using the
Lempel and Ziv algorithm \cite{LZ77} -- available through the
{\tt gzip} program. In this way they obtained an estimation
of the entropy per character $h$, as the ratio between the length of the zipped file and the original length file, when the text length
tends to infinity. They used this method to estimate the entropy per character of the ergodic source  for different values of the flipping
rate, with the result that there exists a close relation between
the current in the ratchet and the net transfer of information in
the system.

The aim of this paper is to apply this technique to a discrete--time and space
version of the Brownian ratchet, known in the literature as Parrondo's paradox.

Parrondo's paradox \cite{HA99a,HA99b,HAT00,HATP01,HA02,brkh99} is based on a
combination of two negatively biased games -- losing games --
which when combined give rise to a positively biased game --
winning game. This paradox appeared as a translation of the
physical model of the Brownian ratchet into game--theoretic terms.
But although it was inspired by the flashing ratchet model, there
has been no quantitative demonstration of their relation until
very recently \cite{TAM03a,TAM03b,AA02}.

More precisely, the so called Parrondo's paradox is based on the
combination of two games (Parrondo's games). One of them, say game A, is a simple
coin tossing game where the player has a probability $p$ of
winning one unit of capital and a probability $1-p$ of losing one
unit of capital. The second game, game B, is a capital
dependent game, where the probability of winning depends on the
capital of the player modulo a given number $M$. For the original
games, $M$ is set to three and there are only two possible values
for the winning probabilities, $p_0$ when the capital is a
multiple of three and $p_1$ otherwise. Their numerical values are
\begin{equation}\label{probab_orig}
   \left\{ \begin{array}{cc}
    p=&\frac{1}{2}-\epsilon,\\
    p_0=&\frac{1}{10}-\epsilon,\\
    p_1=&\frac{3}{4}-\epsilon,
    \end{array}\right.
\end{equation}
here $\epsilon$ is nothing but a biasing
parameter that converts games A and B into losing games. When $\epsilon=0$ it can be demonstrated that the fairness condition is fulfilled for both games, that is $\prod_{j=0}^{M-1}p_i=\prod_{j=0}^{M-1}(1-p_i)$. As soon as $\epsilon>0$ the latter condition no longer applies and A and B are both losing games.

The combination game, game AB, is obtained mixing game A and game
B with probability $\gamma$. As a result of this combination we obtain a winning game, even when $\epsilon$ is equal to zero. The corresponding winning
probabilities are given by the following expressions
\begin{equation}\label{probab_mix}
   \left\{ \begin{array}{cc}
    p'_0=&\gamma p+(1-\gamma) p_0,\\
    p'_1=&\gamma p+(1-\gamma) p_1,
    \end{array}\right.
\end{equation}

Several other versions of the games have been introduced. In the so-called cooperative games\cite{T01,T02}, one considers an ensemble of interacting players. In the history dependent games\cite{PHA00,KJ02}, the probabilities of winning depend on the history of previous results of winnings and losing. Finally, in the games with self--transition\cite{AATA03}, there are non null probabilities that the capital can stay the same (not winning or losing) in a given tossing of the coins.

Some previous works in the literature have related Parrondo's games and
information theory. Pearce, in Ref.\cite{P00}, considers the relation between
the entropy and the fairness of the games, and the region of the parameter
space where the entropy of game A is greater than that of B and AB. Harmer
\emph{et. al} \cite{HATPP00} studied the relation between the fairness of games
A and B and the entropy rates considering two approaches. The first one
consisted on calculating the entropy rates not taking into account the
correlations present on game B, finding a good agreement between the region of
maximum entropy rates and the region of fairness. In the second approach they
introduced these correlations obtaining lower entropy rates and no significant
relation between fairness and entropy rates for game B.

In this paper we aim to relate the current or gain in Parrondo's games with the
variation of information entropy of the binary file generated. Following the
same approach as in Ref.\cite{ASF03}, we have carried out simulations for
different versions of Parrondo's games, including the original version, the
history dependent games, Parrondo's games with self--transition and cooperative
games mentioned above.

\section{Simulation results}

We have carried out numerical simulations of the different versions of the
games. In the simulations, the evolution of the capital of the player has been
converted to a binary string of $0$'s and $1$'s, corresponding to a decrease
and an increase of capital respectively. We must remark that each one of these
characters is stored as a byte (eight bits). A binary file has then been
created for different values of the $\gamma$ parameter. In order to obtain the
desired effect there must be a certain delay time, which we will denote by
$\delta_t$, between measurements of the capital of the player. 

The binary files created have been zipped using the {\tt gzip} (v. 1.3) program, to obtain an estimation of the entropy per character \textit{h}. For a better comparison with the current -- or gain -- generated when alternating between games A and B, we will define $R=\frac{1}{h}$, accounting, in some sense, for the known information about the system. In the Appendix there is an extract of the C program used for the case of the original Parrondo's games.
   \begin{figure}
   \begin{center}
   \begin{tabular}{c}
   \includegraphics[width=10cm]{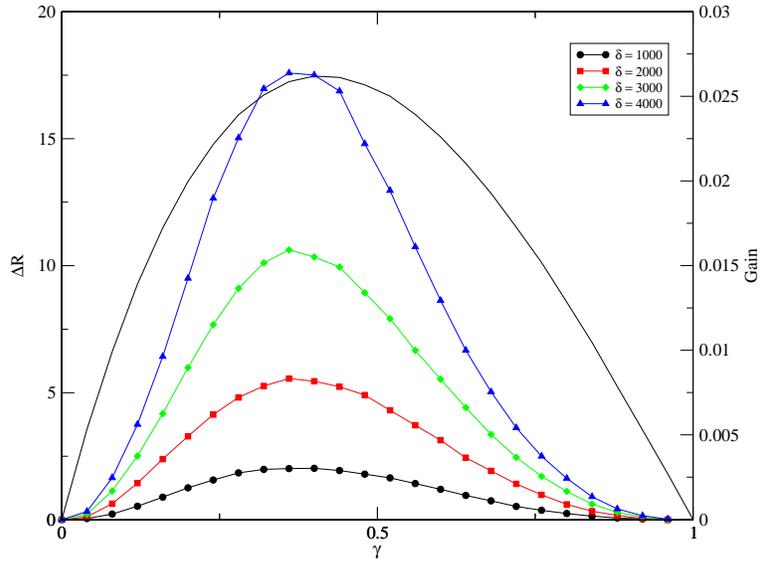}
   \end{tabular}
   \end{center}
\caption{\label{Fig1} Comparison of the gain in the original Parrondo's games with $\Delta R$ calculated for different $\delta_t$ and $\gamma$ values. The probabilities for the Parrondo game are  $p=\frac{1}{2}$, $p_0=\frac{1}{10}$ and $p_1=\frac{3}{4}$. The values used are $\delta_t=1000$ (circles), $2000$ (squares), $3000$ (diamonds) and $4000$ (triangles).}
\end{figure}

   \begin{figure}
   \begin{center}
   \begin{tabular}{c}
   \includegraphics[width=10cm]{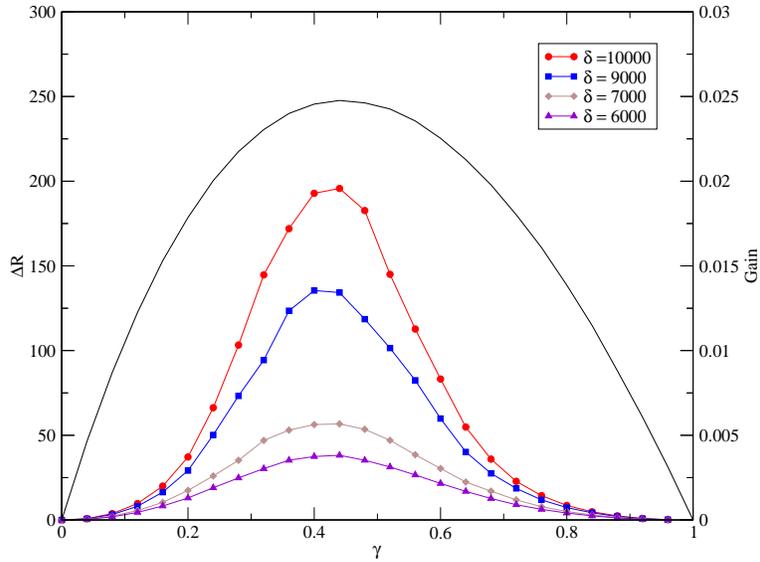}
   \end{tabular}
   \end{center}
\caption{\label{Fig2}Comparison between the inverse of the entropy per character $R$, calculated for different $\delta_t$ values, and the gain for the Parrondo's games $A$ and $B$ with self--transition varying $\gamma$. The values for the probabilities are: $p=\frac{9}{20}$, $r=\frac{1}{10}$, $p_0=\frac{3}{25}$, $r_0=\frac{2}{5}$, $p_1=\frac{3}{5}$ and $r_1=\frac{1}{10}$. See Ref.\cite{AATA03} for an explanation of these parameters. The values used are $\delta_t=6000$ (triangles), $7000$ (diamonds), $9000$ (squares) and $10000$ (circles).}
\end{figure}

In Fig.\ref{Fig1} we compare the gain when alternating between games A and B with probability $\gamma$ with the value of $R$ obtained for different values of the $\delta_t$ parameter. The numerical curves are shifted to the origin, that is, we plot $\Delta R=R_{\gamma}-R_{\gamma=0}$.

We find a good qualitative agreement between the increase in the
gain and the decrease (increase) in entropy ($R$) as the $\gamma$ parameter is varied. Although both curves do not fully coincide, the important
fact to consider here is that the $\gamma$ value for which there
is the maximum decrease in entropy agrees with the value for the
maximum gain in the games. This decrease in the entropy of the
system implies that there exists an increase in the amount of
information that we know about the system.

   \begin{figure}
   \begin{center}
   \begin{tabular}{c}
   \includegraphics[width=10cm]{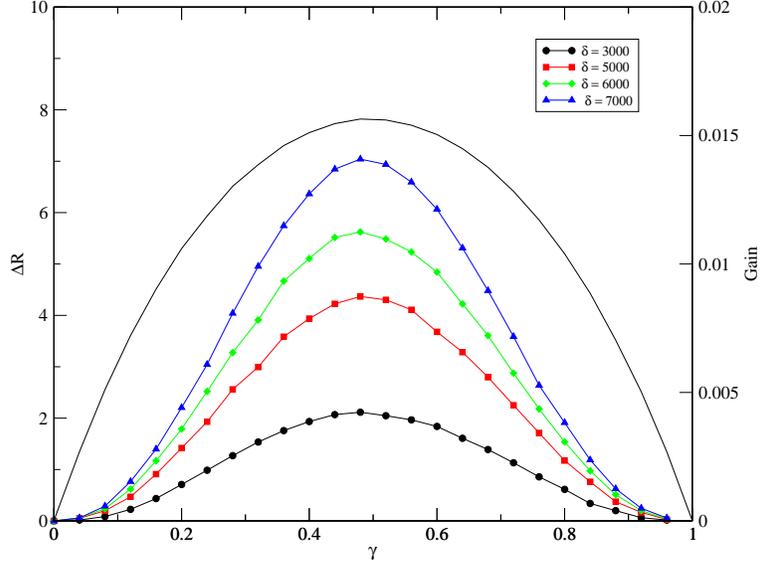}
   \end{tabular}
   \end{center}
\caption{\label{Fig3}Curve for the inverse of the entropy per
character $R$ and the gain for the history dependent games, alternating
between two games with probabilities: $p_1=\frac{9}{10}$,
$p_2=p_3=\frac{1}{4}$, $p_4=\frac{7}{10}$; $q_1=\frac{2}{5}$,
$q_2=q_3=\frac{3}{5}$ and $q_4=\frac{2}{5}$. See Ref.\cite{PHA00} for an explanation of these parameters. The values for the $\delta_t$ are: $3000$ (circles), $5000$ (squares), $6000$ (diamonds) and $7000$ (triangles).}
\end{figure}

In Fig. \ref{Fig2} we compare the gain curve and the entropy curve for the case
of Parrondo's games with self--transition.  Again in this case the maximum gain
coincides with the $\gamma$ value for the minimum entropy per character for all
values of $\delta_t$.

   \begin{figure}
   \begin{center}
   \begin{tabular}{c}
   \includegraphics[width=10cm]{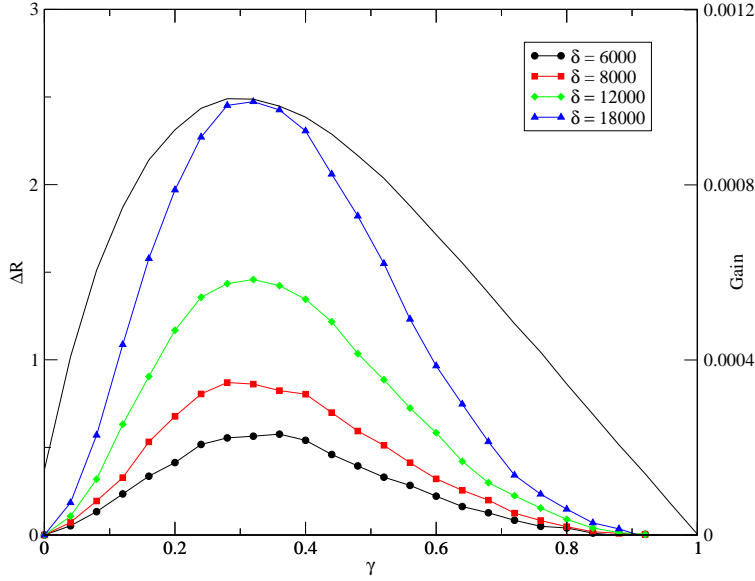}
   \end{tabular}
   \end{center}
\caption{\label{Fig4}Curve for the inverse of the entropy per
character $R$ and the gain for the cooperative version with
probabilities:  $p=\frac{1}{2}$, $p_1=1$,
$p_2=p_3=\frac{16}{100}$, $p_4=\frac{7}{10}$ and $N=150$ players
for different values of the $\gamma$ parameter. See Ref.\cite{T01} for an explanation of these parameters. The values of the $\delta_t$ are: $6000$ (circles), $8000$ (squares), $12000$ (diamonds) and $18000$ (triangles).}
\end{figure}

Finally, in Figs. \ref{Fig3} and \ref{Fig4} we show the comparison
in the case of the history dependent games, and
cooperative games, sharing all them the same features as
the previous games analyzed before.

\section{Theoretical analysis}

In this section we develop a simple argument that explains the observed
relation between the gain of the games and their entropy. In the seminal work
by Shannon\cite{S48} the entropy per character of a text produced by an ergodic
source is given by the following expression

\begin{equation}\label{entropia}
H=-\sum_ip_i\cdot \log(p_i)
\end{equation}

\noindent where $p_i$ denotes the probability that the source will `emit' a
given symbol $a_i$, and the sum is taken over all possible symbols
that the source can emit.

If the source can be found in several states, each one with
probability $P_j$, then the entropy reads
\begin{equation}\label{suma_entropia}
H=\sum_{j=1}^{M} P_j\,H_j=-\sum_{j,i}P_j\,p_i^j\,\log(p_i^j).
\end{equation}
In the last equation $p_i^j$ denotes the probability of emitting
the symbol $a_i$ when the source is found in the state $j$.

Remember that we measure and store the capital of the player
after $\delta_t$ rounds have been   played. When $\delta_t$ equals one it means that we
measure the capital of the player \textit{in each} round.
Translating this into entropic terms we have that, considering the
original Parrondo's games, our source will be in two possible
states, say, when the capital is multiple of three and otherwise.
From Markov chain theory \cite{KT75} we know that the stationary
probabilities of finding the capital in each of the states are
given by

\begin{equation}\label{stat_prob}
\begin{array}{cc}
     \Pi_0=\frac{1}{D}(1-p_1+p^2_1)\\
     \\
     \Pi_1=1-\Pi_0=\frac{1}{D}(2-p_0-p_1+2\,p_0p_1)
\end{array}
\end{equation}
where $D=3-p_0-2p_1+2p_0p_1+p^2_1$.

It is interesting to consider our source (the capital of the player) as
composed of several sources, each one with an homogeneous statistical structure, i.e. they are ergodic sources. Each source then will have a determined
probability of `emitting' a zero or a one, as the player has a different
winning probability depending on whether his capital is multiple of three or
not. If we consider that the source will emit the symbol $1$ with
probability $p_i$ and $0$ with probability $1-p_i$ then Eq.
\ref{suma_entropia} reads

\begin{equation}\label{suma_entropia_a}
\begin{array}{ll}
H&=\sum_{j=1}^{M}\Pi_j\,H_j=-\sum_{j,i}\Pi_j\,p_i^j\,\log(p_i^j)\\
&=-\Pi_0\cdot \left(p_1\cdot \log(p_1)+(1-p_1)\cdot
\log(1-p_1)\right)-\left(1-\Pi_0\right)\cdot \left(p_2\cdot
\log(p_2)+(1-p_2)\cdot \log(1-p_2)\right)
\end{array}
\end{equation}

\begin{figure}
   \begin{center}
   \begin{tabular}{c}
   \includegraphics[width=10cm]{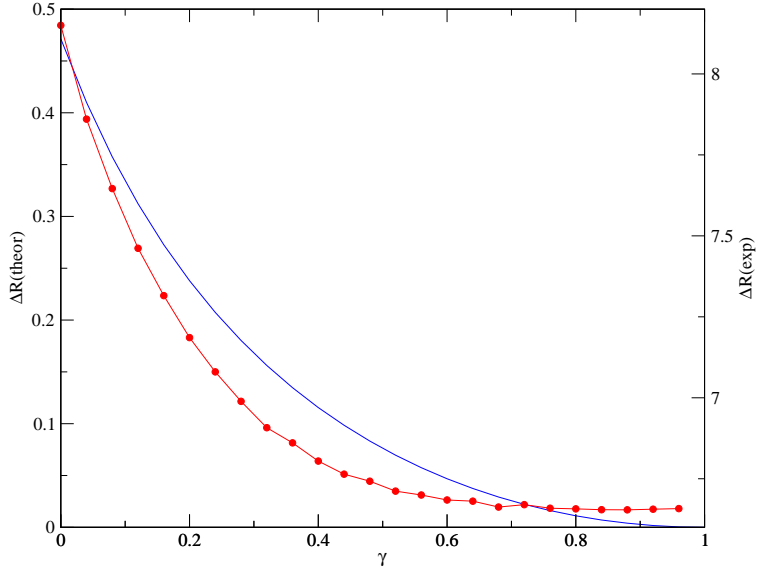}
   \end{tabular}
   \end{center}
\caption{\label{Fig5} Plot of the theoretical curve for the inverse of the 
entropy per character and the numerical curve (circles) for the case when
$\delta_t=1$ step.}
\end{figure}

In Fig.\ref{Fig5} we plot the inverse of Eq.\ref{suma_entropia_a} for the values
of $p_i^j$ and $\Pi_i$ corresponding to the original Parrondo's games together with the numerical curve obtained when zipping the files obtained with $\delta_t=1$. It is worth noting how the negentropy decreases as $\gamma$ increases, translating into an increase of known information about the system.

Alternatively, we can now estimate how the entropy changes for
$\delta_t\gg 1$, when we allow a large amount of time
between measures. As $\delta_t$ is increased, the statistical
description used above cannot be used anymore. For higher $\delta_t$
values, the system gradually loses its memory about its previous
state and so what we obtain are measures that can be considered as
statistically independent, as we will not know for sure which will
be the probability of finding the \emph{source} in any of the two
possible states. Now the dominant term will be the averaged
probability --over all possible states-- of finding the capital
of the player with a value higher than the previous measure.

Considering the same stationary probabilities $\Pi_0$ and $\Pi_1$
given by Eq.\ref{stat_prob} we can obtain the expressions for the
average winning and losing probabilities over these two states as

\begin{equation}
\begin{array}{cc}
p_{win}&=\sum_{j=1}^{M} \Pi_j\,p^i_j=\Pi_0\,p'_0+\Pi_1\,p'_1 \label{pwin}\\
p_{lose}&=\Pi_0\cdot(1-p'_0)+\Pi_1\cdot(1-p'_1)\label{plose}
\end{array}
\end{equation}

It can be noticed in all Figs.\ref{Fig1},\ref{Fig2},\ref{Fig3},\ref{Fig4} that as $\delta_t$ increases, the entropy curve acquires a more pronounced peak. This effect can be explained due to the fact that as the time between measures increases, the probability of finding the player with a higher (lower) capital
increases (decreases). We can give an estimate of the probability that after $\delta_t$ rounds played, the capital of the player will be greater ($p_>$) than it was initially by considering the probability that there have been less that $\delta_t/2$ loses in the $\delta_t$ coin tosses,

\begin{equation}
p_{_>}=\sum_{k=0}^{\frac{\delta}{2}}
{\delta_t \choose k}\cdot p_{win}^{\delta_t-k}\cdot p_{lose}^k
\end{equation}

\begin{figure}
   \begin{center}
   \begin{tabular}{c}
   \includegraphics[width=10cm]{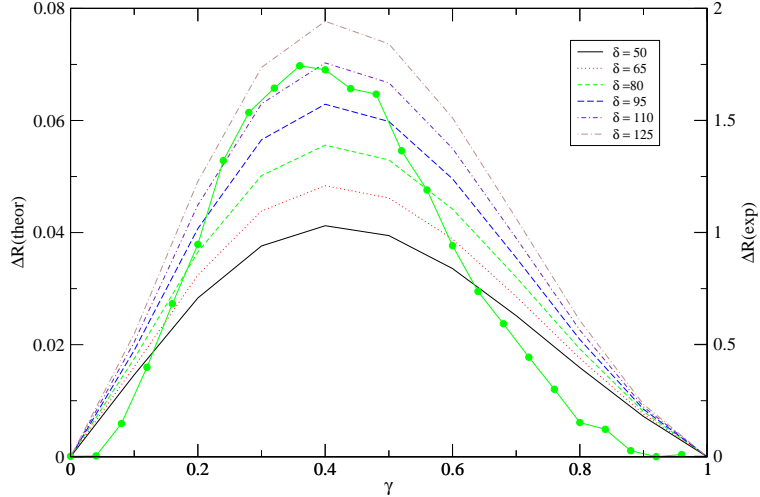}
   \end{tabular}
   \end{center}
\caption{\label{Fig7} Plot of the theoretical curves for the inverse of the
entropy per character for different values of $\delta_t$, together with the numerical curve obtained for $\delta_t=500$ steps.}
\end{figure}

These previous equations together with 
\begin{equation}\label{entropia_win}
H=-p_{_>}\cdot \log(p_{_>})-(1-p_{>})\cdot \log(1-p_{_>}),
\end{equation}
allows one to obtain the entropy per character. In Fig.\ref{Fig7} we can see the curve for the variation of the inverse of information entropy for different values $\delta_t$. In this case the negentropy decreases as $\delta_t$ increases, reflecting an increase of known information about the system.

This theoretical explanation also holds for the case of  history dependent Parrondo's games and Parrondo's games with self--transition. For the latter case, though, we must take into account the fact that there exists a non negligible possibility that the capital can remain the same after a round played. In this case we need only to consider a source that will emit three different symbols, depending on whether the capital is greater, equal or smaller than our previous measure. 

As a conclusion, we have demonstrated that there exists a transfer
of information in the case of Parrondo's games, considered as a
discrete--time and space version of the flashing ratchet. This
effect takes place in every existing version of the games
analyzed, showing the robustness of it. The qualitative analysis
for the case of the original Parrondo's games have demonstrated
the fact that the capital of the player can be thought of as a mixed
source, composed of two ergodic sources, depending on whether the
capital is a multiple of three or not. 

We have also given some insight about the behavior of the information entropy for low and high values of $\delta_t$, and why there is a decrease (increase) in the entropy per character $h$ (information) as $\delta_t$ increases. 

\section*{Acknowledgments}
We thank C. M. Arizmendi for several discussions, including the sharing of useful information about the method. We acknowledge financial support from
MCYT (Spain) and FEDER through projects BFM2000-1108 and
BFM2001-0341-C02-01. P.A. is supported by a grant from the Govern Balear.

\section*{APPENDIX}

We provide the C program used for creating the binary files that will be zipped for obtaining the entropy per character \emph{h}.

\begin{verbatim}
int main()
{
    int i, j, k, ijk;
    int ijk1, ij, n;
    long nmax, irepet, icont, deltat, t;
    long  Nrepet, c, c0;
    double gamma, pa, pb1, pb2;
    double u, v, gain;
    unsigned char cero=0;
    unsigned char uno=255;

    srand48(342342);
    
    
    FILE *current;
    current = fopen("gain.dat","w");

/* We fix the lenght in bytes of the binary file   */

    Nrepet=300000;

/*  We impose the delay time between measurements  */ 

    deltat = 500;


/*  Probabilities of the original Parrondo's games */

    pa = 1./2.;
    pb1 = 1./10.;
    pb2 = 3./4.;


    ijk = 0;

    for (gamma = 0.; gamma <= 1.0; gamma = gamma + 0.04){

/*  we create a binary file for each value of gamma  */ 

    sprintf(q,"gamma%d.dat",ijk);
    FILE *data;
    data = fopen(q,"w");
    ++ijk;
 
    c0 = 0;
    c = 0;
    i= 0;
    ijk1 = 0 ;
    t = 0;
        
	while (i<Nrepet){
	    ++t;
        if (drand48() < gamma ){

/*      we  play game A    */

            if (drand48() < pa ){
            ++c;}
            else {
            --c;
            }
        ++ijk1;}
        else{

/*      we play game B   */

            if (fmod(c,3) == 0 ){
                if (drand48() < pb1 ){
                ++c;}
                else {
                --c;}
            }
            else {
                if (drand48() < pb2 ){
                ++c;}
                else {
                --c;}
            }
        ++ijk1;}

/* 	every time that ijk1 equals deltat we evaluate the actual capital 
	of the player compared to its previous value 		  */

        if (ijk1 == deltat) {
        ijk1 = 0;
            if ((c-c0) > 0) {
            putc(cero,data);}
            else{
            putc(uno,data);
            }
        i++;
        c0 = c;
        }
        }
    
/*	we also write the gain for every gamma value		*/

    gain =(double) c/((double) t);
    fprintf(current,"%f\t%f\n",gamma,gain);
    fclose(data);
    }
fclose(current);
return EXIT_SUCCESS; 
}
\end{verbatim}

\end{document}